\documentclass[preprintnumbers,amsmath,amssymb,floatfix,11pt, prd,onecolumn,
superscriptaddress,nofootinbib]{revtex4}
\usepackage{latexsym}
\usepackage{epsfig}
\usepackage{epstopdf}
\usepackage{amssymb}
\begin{document}
\title{\bf  Dynamics and Centre of Mass Energy of Colliding Particles Around Black Hole in $f(R)$ Gravity}
\author{Bushra Majeed}
\email{bushra.majeed@sns.nust.edu.pk}\affiliation{School of Natural
Sciences (SNS), National University of Sciences and Technology
(NUST), H-12, Islamabad, Pakistan}
\author{Mubasher Jamil}
\email{mjamil@sns.nust.edu.pk}\affiliation{School of Natural Sciences (SNS),
National University of Sciences and Technology (NUST), H-12,
Islamabad, Pakistan}
\begin{abstract}
{\bf Abstract:} We have investigated the dynamics of particles in the vicinity of a static spherically symmetric black hole in $f(R)$ gravity. Using the Euler Lagrange method the dynamical equations of a neutral particle are obtained. Assuming that the particle is initially moving in the innermost stable circular orbit, we have calculate its escape velocity, after a collision with some other particle. The conditions for the escape of colliding particles are discussed. The effective potential and the trajectories of the escaping  particles are studied graphically.
\end{abstract}
\maketitle
\newpage
\section{Introduction}
Observations of Cosmic Microwave Background (CMB) and Supernova type Ia (SN Ia) \cite{f1, f2} indicate that our universe is in the phase of accelerated expansion. There are some modifications in the general relativity which  explain this expansion phenomena. Dark energy is one of the well known proposals for explaining the expansion behavior, but the modifications coming from the vacuum energy in quantum field theory, result in a huge amount energy, much larger than the required for observed expansion rate. It is contended that these modification  do not obey the weak equivalence principle \cite{f4, f6}.
One of the modified theory is $f(R)$ theory of gravity \cite{f8, f9, f10, f11, f12, f13}, it explains the accelerated expansion of the universe. There are a lot of valid reasons because of which $f(R)$ gravity has became one of  the most interesting theory in this era. It is an extension/ modification in the  Einstein's general relativity. The $R$ stands for the Ricci
scalar and  $f(R)$ is an analytic function of $R$ (for review see \cite{nojri11507, sotiriou45110, nojri5911}).
 It is one of the most studied modified theory of gravity, having astrophysical consequences \cite{f19, f20}. It is  important
and interesting  to study the astrophysical phenomenon
 in $f(R)$ gravity.

 Dynamics of particles around black holes in the theory of relativity has been studied widely.
  %Geodesics may display a rich structure and they can
%convey a very reliable information to understand the geometry of the
%BH.
 Motion of test
particles give information about the gravitational fields of the objects
experimentally and  the observations are compared with the predictions \cite{frolov98, sh5979, podurets64, ames68}.
 Presence of magnetic field also effects the motion of charged particles around the black holes \cite{zn7076, bl3377}. It transfers
energy to the moving particles, in the geometry around black holes, and helps them to escape to spatial infinity \cite{ko8802, mi4199, te0903, saqib, ja2415}. Weakly magnetized black holes have magnetic field around them \cite{fr2012}, it is because of plasma around the black holes \cite{bo2013, mc2307, do2008}. Geometry of the black hole is not effected by a weak magnetic field, with strength $B\sim 10^4-10^8\ll 10^{19}$ Gauss. But it has considerable effects near horizons of the black holes.

The geometrical structure of  the spacetime around a black hole  could be understood better by studying the dynamics of particles in its vicinity \cite{frolov1998, sh5979}.
In literature many aspects of particles motion in the vicinity
of black holes have been studied. In \cite{gulmina} dynamics of  charged particles in the geometry of weakly magnetized naked singularity, Janis-Newman-Winicour (JNW) spacetime, has been studied. Dynamics of
charged particles in the vicinity of a magnetized five dimensional rotating black hole has been studied by Kaya \cite{ka1107}. Dynamics of particles around  magnetized black holes in the modified gravity are studied in \cite{saqib-mog}. Ba˜nados, Silk and West (BSW) proposed that rotating black holes may act as particle
accelerators \cite{BSW}. Dynamics of particles and the phenomena of high energy collision of particles in the centre of mass energy frame, is studied in \cite{za7110}. In \cite{pu5211} the spatial regions for the circular motion of the neutral and
charged test particles around the Reissner-Nordstr\"{o}m black hole and the naked singularities have been studied.
%Plyatsko and Fenyk estimated
%the motion of spinning test particles in planar circular orbits
%around Kerr BH \cite{fen}.
The BSW effect in the vicinity of slowly rotating black hole in Horava-Lifshitz gravity and a 3+1 dimensional
topological black hole has been studied \cite{ibrar}.
Critical escape velocity of charged and neutral particles moving around the weakly magnetized Schwarzschild black hole \cite{za4313}, slowly rotating Kerr black hole immersed in the magnetic field \cite{saqib-kerr}, the weakly magnetized Reissner-Nordstr\"{o}m black hole \cite{majeedRN} and in the vicinity of a Schwarzschild-like black hole
in the presence of quintessence and magnetic field \cite{jamilschwarz} have been studied.   %In \cite{hu0815} timelike geodesics for modified gravity black hole immersed in an axially symmetric magnetic field has been studied.

 In this paper we have discussed the dynamics of particles in the geometry of a static, spherically symmetric black hole in $f(R)$ theory of gravity. The phenomenon of collision between the particles in the vicinity of the $f(R)$-BH, following the approach of Zahrani $\text{et. al.}$ \cite{za4313} has been studied. The circumstances under which the particle can escape to
infinity after collision are studied.
The order of the paper is as follows: In section II we discuss the metric for a static spherically symmetric black
hole in a particular model of $f(R)$ gravity and the escape
velocity of a neutral particle is obtained. Behavior of effective potential and trajectories of the escaping particles are plotted in section III. In section IV trajectories for the escape energy and the escape
velocity of the particle are presented. For simplicity we restrict these calculations for the particle initially moving in the equatorial plane. The metric signature is $(-,+,+,+)$ and
$c=1,G=1$.

\section{Dynamics of a Neutral Particle in $\textbf{f(R)}$ Spacetime}
We discuss the motion of a neutral particle near  black hole in $f(R)$ gravity. Metric of the $f(R)$-BH is \cite{f17}
\begin{eqnarray}\label{f1}
ds^{2}&=& -f(r)dt^2+\frac{1}{f(r)}dr^2 +r^2(d\theta^2 +\sin^2\theta
d\phi^2),
\end{eqnarray}
where
\begin{equation}
\label{metric}f(r)=1-\frac{2M}{r}+\beta r-\frac{\Lambda r^2}{3},
\end{equation}
here $M$ is the mass  and $Q$ is electric charge of the black hole, $\beta=\alpha/d\geq 0$  is a constant \cite{f17}, with $d$ as a scale factor and $\alpha$ is a dimensionless parameter. %$ \approx 10^{-6}$.
Dynamics of the particle can be studied by using the Euler Lagrange equations \begin{equation}m\frac{D u^{\mu}}{d\tau}=0, \end{equation}where $D/d\tau$ represents the covariant derivative of the metric and $u^{\mu}$ is the four velocity of the particle, with normalization condition $u^{\mu}u_{\mu}=-1$.
There are two symmetries of the black hole metric along the time translation and rotation around symmetry axis. The corresponding conserved quantities (total energy $\mathcal{E}$ and specific azimuthal angular momentum $L_{z}$) can be obtained from the Killing vectors
\begin{equation}
\xi_{(t)}^{\mu}\partial_{\mu}=\partial_{t} , \qquad
\xi_{(\phi)}^{\mu}\partial_{\mu}=\partial_{\phi},
\end{equation}
here
$\xi_{(t)}^{\mu}=(1,~0,~0,~0)$ and $\xi_{(\phi)}^{\mu}=(0,~0,~0,~1)$. The energy of the particle is
\begin{equation}\label{f2}
  \mathcal{E}=-p_{\mu}\xi^{\mu}_{(t)}/m=\dot{t}f(r),
\end{equation}
\begin{equation}\label{f3}
  L_{z}=p_{\mu}\xi^{\mu}_{(\phi)}/m=\dot{\phi}{r^2 \sin^2 \theta},
\end{equation}
here dot denotes the differentiation with respect to proper time
$\tau$. Specific total angular momentum of the particle is \begin{equation}L^2=r^4 \dot{\theta}^2+\frac{L_z^2}{\sin^2 \theta}=r^2 v_{\bot}^2+\frac{L_z^2}{\sin^2\theta},\end{equation} here $v_{\bot}\equiv-r \dot{\theta_o},$ and $\dot{\theta}_o$ denotes the initial angular velocity of the  particle. From the
normalization condition $u^{\mu}u_{\mu}=-1$, the $r$ component of velocity vector is
\begin{equation}\label{f4}
\dot{r}^{2}=\mathcal{E}^2- U_\text{eff},\end{equation}here
\begin{equation}
V_{\text{eff}}= (1 - \frac{2 M}{r} +\beta r- \frac{\Lambda r^2}{3})(1 +
\frac{L_z^2}{r^2}),
\end{equation} % since
since $r>0$ and real so for the physically suitable regions $r$ should satisfy ${\mathcal{E}^2} \geq V_{\text{eff}}$. The metric is spherically symmetric so we consider the motion of the particle in the equatorial plane, $\theta=\pi/2$. For the planer motion chose $\dot{\theta}=0$ and the effective potential becomes \begin{equation}V_{\text{eff}}=\tilde{U}\equiv f(r)(1+\frac{L_z^2}{r^2}).\end{equation}
Now we investigate the properties of the effective potential. For the motion of a particle in the exterior of black hole, $r$ must always be greater than zero. The minimum distance from the black hole at which a particle orbits around it is known as the innermost stable circular orbit (IMSCO), $r_o$, which could be obtained by solving $V_{\text{eff}}$ for its extreme values, i.e. the root of $dV_{\text{eff}}/dr=0$. The equations which determine the IMSCO are
\begin{equation}\frac{dV_{\text{eff}}}{dr}=\frac{6Mr^2+3L_z^2(6M-r(2+r\beta))+r^4(3\beta-2 r \Lambda)}{3 r^4}, \end{equation} and \begin{equation}\frac{d^2V_{\text{eff}}}{dr^2}=-\frac{2(6Mr^2+L_z^2 (36M-3r(3+r\beta))+r^5\Lambda)}{3 r^5}. \end{equation}
Energy and the azimuthal
angular momentum in the IMSCO are
\begin{equation}\label{fee}
  \mathcal{E}_{o}=\frac{\sqrt{2}\Big(6M+r_o(-3-3r_o\beta+r_o^2\Lambda)\Big)}{\sqrt{9r_o(-6M+r_o(2+r_o\beta))}},
\end{equation}
and
\begin{equation}\label{cirf1} L_{zo}=\pm \frac{\sqrt{-6Mr_o^2-3r_o^4\beta+2r_o^5\Lambda}}{3(6M-2r_o-r_o^2 \beta)}.
\end{equation}
Depending on the energy $\mathcal{E}$ and the angular momentum $L_z$ of the
particle, one can determine nature of its orbit around the black hole, it also depends on the cosmological constant $\Lambda$ and value of
constant $\beta$. Consider that the particle initially moving in the IMSCO collides with a particle, coming from infinity (initially at
rest). The collision may cause the particle to have the following possible outcomes of its motion: (i) a bounded orbit (ii) captured
by the black hole (iii) escape to infinity. Collision of the particles changes the equatorial
plane of the moving particle but since the metric is spherical
symmetric so  all the equatorial planes are similar.  We assume that  the collision  occurs  in such a manner  that a small change occurs only in their energies while
the azimuthal angular momentum and the initial radial
velocity do not change. So the orbit of the particle refines slightly and only the change in energy will be counted for determining the motion of the particle after collision.
These condition are imposed because otherwise the particle would move away from the original
path which will cause the particle to be captured by the black hole or escape to infinity. Particles gains an escape velocity  $(v_{\text{esc}})=v_{\bot}$ in
the orthogonal direction of the equatorial plane after collision
\cite{frolove3410} and its momentum and energy (in the new  equatorial plane) are
\begin{equation}
  L^{2}=r_{o}^{2}V_\text{esc}^{2}+L_{zo}^{2},
\end{equation} here $V_\text{esc}\equiv -r\dot{\theta}_{o}$. Further
\begin{equation}\label{f33}
 \mathcal{E}_\text{new}=\sqrt
{\frac{(\Lambda r^3-3\beta r^2+6M-3r)}{3r}V_\text{esc}^2 + \mathcal{E}^2_o},
\end{equation} with  $\mathcal{E}_{o}$, given in Eq. (\ref{fee}).

After collision, the particle gains greater angular momentum and energy as compared to before collision.
From Eq. (\ref{f33})  it is clear that in the asymptotic limit
($r\rightarrow\infty$),
$\mathcal{E}_\text{new}\rightarrow\mathcal{E}_{o}= 1$. So for unbounded motion (escape) particle requires
$\mathcal{E}\equiv \mathcal{E}_\text{new} \geq1$ .  Hence for escape
to infinity the necessary condition is %$\mathcal{E}_{new}\geq1$\\or
\begin{equation}
V_\text{esc}\geq \frac{\sqrt{3(r-r\mathcal{E}_o)}}{\sqrt{6M-3r-3r^2\beta+\Lambda r^3}},
\end{equation}
 we have solved equation $(\ref{f33})$ taking $\mathcal{E}_\text{new}\ge1$.
\begin{figure}
\centering
\includegraphics[width=10.0cm]{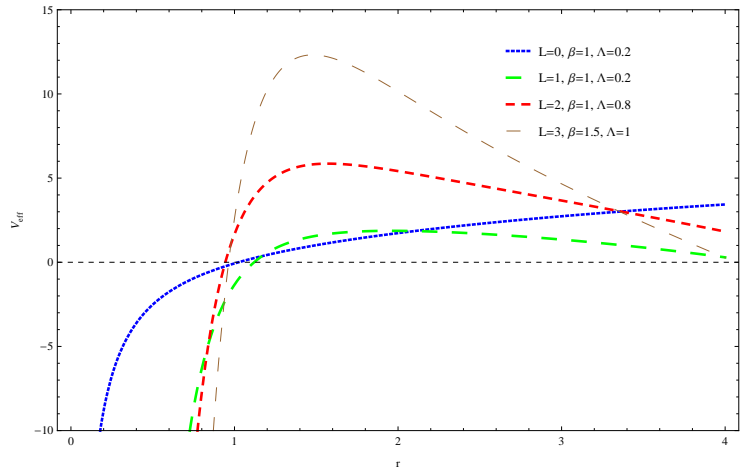}
\caption{Behavior of the effective potential versus $r$, for different values of the parameters.}\label{veff1}
\end{figure}
\section{Behavior of the Effective Potential and Trajectories of Escape Velocity}
Effective potential of the moving particle help us to analyze its
stable and unstable circular orbits. Dynamics of the particle,
given by $\dot{r}^2=E^2- V_\text{eff}^2$  is possible only if  $E^2-\geq V_\text{eff}^2$.
For the particle, moving radially towards the black hole, the
geodesics can be obtained by using (\ref{f2}) and (\ref{f4}) together
with zero angular momentum, $L=0$,
\begin{equation}
\label{1a}\frac{d r}{d t}\equiv \frac{\dot{r}}{\dot{t}}
= \pm \frac{\sqrt{\mathcal{E}^2-f(r)}}{\mathcal{E}f^{-1}(r)}.
\end{equation}
A better understanding of the geodesics can be obtained by ploting the geodesics equation in ($t,r$)-coordinates \cite{ibrar} . Since the metric contains the horizons, (coordinate singularities), these singularities do not allow us to find the full ploting of the geodesics so one need to move in some non-sigular coordinates system as done in \cite{azad, bushra}.
The particle can have circular orbit at those $r$ where
$V_\text{eff}$ attains it maximum and minimum values. The circular orbits are
stable at the position where $V_\text{eff}$ is minimum and for the maximum value of $V_\text{eff}$ the circular
orbits are unstable. At $\frac{d^2V}{dr^2}=0$ ,the point of inflection i.e. the marginally stable orbit occur.

\begin{figure}
\centering
\includegraphics[width=10.0cm]{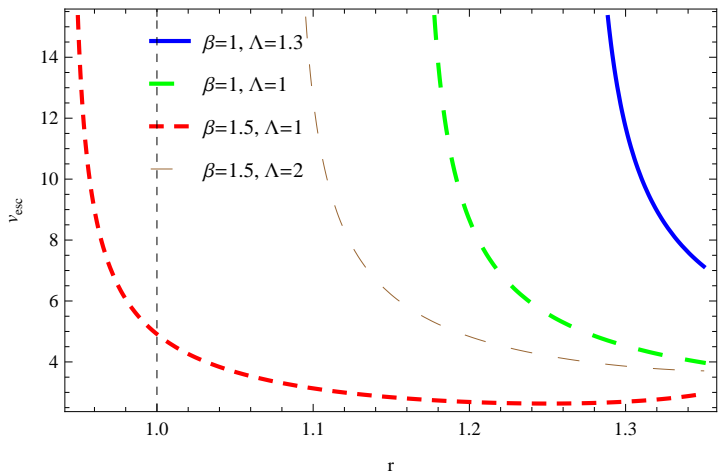}
\caption{Escape velocity of the particle versus $r$, for different values of the metric parameters.}\label{esc1}
\end{figure}

In Fig. (\ref{veff1}) a comparison of the effective potential for different values of the parameters $L$, $\beta$ and $\Lambda$ is plotted. It is observed that for the larger values of $L$, the particle has high effective potential. The effective potential has
 relative maxima in the surrounding of black hole horrizon, it corresponds to the unstable orbits.
The escape velocity is plotted in Fig. (\ref{esc1}). It is clear from the figure  that as values of the metric parameters, $\Lambda$ and $\beta$, change the behavior of the velocity of the particle, required to escape, also changes. It shows that the particles can have a maximum or minimum escape velocity depending upon the nature/value  of these parameters.

%\begin{figure}
%\includegraphics[width=10cm]{BSW11.eps}
%\caption{Geodesics for a radially ingoing particle coming from
%infinity with some initial velocity, reaching $r=r_b$ (dashed line),
%and going back to infinity. We chose $\alpha=1$,
%$\mathcal{E}^2=1.5$.} \label{BSW11}
%\end{figure}

%\begin{figure}
%\includegraphics[width=10cm]{lif_veff.eps}
%\caption{Evolution of effective potential versus radius $r$,
%parameters values are set as, $L^2=3$, $\alpha=1$, $m=1$.  At
%$r=r_o=3$ there is a circular orbit of the particle. \label{Veff1}}
%\end{figure}

 \section{Center of Mass Energy of the Colliding Particles}
The collision energy of two neutral particles of masses, $m_1$ and
$m_2$ falling freely coming  from infinity, initially at rest is studied in this section. Consider  the particles are in the equatorial plane and collide near the black hole at some distance $r$ from the event horizon of the black hole. Since the $i$-th particle has $4$-momentum as
 \begin{equation}P^{\mu}_i=m_i u^{\mu}_i~~~~i=1, 2,\end{equation} where $u^{\mu}_i$, $P^{\mu}_i$ and $m_i$ are the $4$-velocity, $4$-momentum and rest mass of the $i$th particle.
 The center of mass frame for some particle is the one in which its total $4$-momentum has zero spatial
component. For the colliding particles, the energy measured by an observer who
is at rest with respect to the center of mass frame is called the center of mass energy
(CME). The phenomenon of arbitrary high CME is a purely relativistic phenomenon and
serves as an excellent tool to study the high energy astrophysics. Let us consider that two
neutral particles collide in the surrounding of the black hole, the collision energy of the particles ($m_1=m_2=m_o$) in the center of mass frame is
 is defined as \cite{bsw}
\begin{equation}\label{ecm} E_{cm}=m_o\sqrt{2}\sqrt{1-g_{\mu\nu} u_1^{\mu}
u_2^{\nu}},\end{equation}where  \begin{equation} u^{\mu}_i\equiv
\frac{dx^{\mu}}{d\tau},~~~i=1, ~2
\end{equation}
is the $4$-velocity of each of the particles. Using Eq. (\ref{ecm}) along with Eqs. (\ref{f2}),
(\ref{f3}), and (\ref{f4}), the CME of the
neutral particles comes out,
\begin{eqnarray}\label{aa9}
E_{cm}&=& m_o\sqrt{2} \Big[2+(\frac{L_1^2 + L_2^2}{2
r^2})(\frac{\mathcal{E}^2 + 1-\frac{2M}{r}+\beta r-\frac{\Lambda r^2}{3}}{\mathcal{E}^2})+ \nonumber\\&& (L_1
L_2)\Big(\frac{\Big(1-\frac{2M}{r}+\beta r-\frac{\Lambda r^2}{3}\Big)(L_1 L_2)- 2 r^2 \mathcal{E}^2}{2 \mathcal{E}^2
r^4}\Big)+ \frac{1-\frac{2M}{r}+\beta r-\frac{\Lambda r^2}{3}}{2 \mathcal{E}^2} \Big]^{1/2}.
\end{eqnarray}
CME of the particles near the
 horizons, $r_+$ and $r_-$, can be calculated by taking $1-\frac{2M}{r}+\beta r-\frac{\Lambda r^2}{3}=0$
\begin{equation}\label{a10}
E_{cm}=2m_o\sqrt{1+\frac{1}{4r_h^2}(L_1- L_2)^2},
\end{equation}here $r_h$ represents the horizons. Notice that the CME would become
infinite if one of the particles has
infinite angular momentum, but to reach the black hole the particles must have some finite value. Hence the CME of the colliding particles in the geometry of $f(R)$ black hole, given in Eq. (\ref{a10}), remains finite.
\section{Summary and Conclusion}
We have considered the phenomena of particles collision in the vicinity of a static spherically symmetric black hole in $f(R)$ gravity, written as $f(R)$-BH. When a particle moving in the innermost stable circular orbit (IMSCO) strikes with some other particle, which is initially at rest and coming from infinity, it may get captured by the black hole, or escape from the black hole and go to infinity or keep on orbiting around the black hole. In the later case as well the particle finally gets captured by the black hole. The final state depends on the energy of the colliding particles. We have derived the velocity required for escape of a neutral particle initially moving in the IMSCO when it strikes with some other particle.
Dynamics equations for a neutral particle are discussed, the energy and the azimuthal angular
momentum of the particle are calculated. We find out the conditions on the energy of the particle
required to escape or to remain bounded in orbit. Expression for the
escape velocity of a neutral   particle moving around the $f(R)$-BH, is
derived and the behavior is observed graphically. The Fig. (\ref{esc1}) shows that changing the metric parameters, $\Lambda$ and $\beta$, behavior of the escape velocity also varies, hence a particle can have maximum or minimum escape velocity depending upon these parameters.
Behavior of the effective potential is also studied graphically, the angular momentum of the particle effect the behavior of its effective potantial, greater the angular momentum more unstable orbit of the particle is observed.

 The calculations for the centre of mass energy of the colliding particles show that it remains finite at the horizon of the black hole. Hence an infinite energy can not be obtained from this collision phenomenon in the $f(R)$-BH.
  

\begin{thebibliography}{99}
\bibitem{f1} A. G. Riess, et al., Astron. J. \textbf{116}, 1009 (1998).
\bibitem{f2} J. L. Tonry, et al., Astrophys. J. \textbf{594}, 1 (2003).

\bibitem{f4}S. Weinberg, Rev. Mod. Phys. \textbf{61}, 1 (1989).
%\bibitem{5} T. Padmanabhan, Phys. Rept. 380, 235 (2003)
\bibitem{f6} J. Polchinski, hep-th/0603249
%\bibitem{f7} J. Martin, Comptes Rendus Physique \textbf{13}, 566 (2012).
\bibitem{f8} A. A. Starobinsky, Phys. Lett. B \textbf{91}, 99 (1980).
\bibitem{f9} N. Ohta, R. Percacci and G. P. Vacca, Phys. Rev. D \textbf{92},
061501 (2015).
\bibitem{f10} S. Soroushfar, R. Saffari, J. Kunz and C. Lammerzahl,
Phys. Rev. D \textbf{92}, 044010 (2015).
\bibitem{f11} M. U. Farooq, M. Jamil, D. Momeni, R. Myrzakulov, Can.
J. Phys. \textbf{91}, 703 (2013).
\bibitem{f12} M. R. Setare, M. Jamil, Gen. Relativ. Gravit. \textbf{43}, 293 (2011).
\bibitem{f13} I. Hussain, M. Jamil, F. M. Mahomed, Astrophys. Space
Sci. \textbf{337}, 373 (2012).

\bibitem{nojri11507}S. Nojiri and S. D. Odintsov, Int. J. Geom. Methods Mod. Phys. \textbf{4}, 115 (2007).
\bibitem{sotiriou45110}T. P. Sotiriou and V. Faraoni, Rev. Mod. Phys. \textbf{82}, 451 (2010).
\bibitem{nojri5911} S. Nojiri and S. D. Odintsov, Phys. Rep. \textbf{505}, 59 (2011).
\bibitem{f19}L. Lombriser, F. Simpson and A. Mead, Phys. Rev. Lett.
\textbf{114}, 251101 (2015).

\bibitem{f20} S. Chakraborty, Class. Quantum. Grav. \textbf{31}, 055005 (2014).

\bibitem{frolov98}V. P. Frolov,  I. D. Novikov, Black Hole Physics, Basic Concepts and New Developments, (Springer
1998).

\bibitem{podurets64} M. A. Podurets, Astr. Zh. \textbf{41}, 1090 (1964).
\bibitem{ames68} W. L. Ames, and K. S. Throne, Astrophys. J. 151, 659 (1968).
\bibitem{sh5979}N. A. Sharp, Gen. Rel. Grav. {\bfseries{10}}, 659 (1979).
\bibitem{zn7076} R. Znajek, Nature {\bfseries{262}}, 270 (1976).
\bibitem{bl3377} R. D. Blandford,  R. L. Znajek, Mon. Not. Roy. Astron. Soc. {\bfseries{179}}, 433 (1977).
\bibitem{ko8802}  S. Koide, K. Shibata,  T. Kudoh, and  D. l. Meier, Science {\bfseries{295}}, 1688 (2002).
\bibitem{ibrar}I. Hussain, M. Jamil, B. Majeed,
Int. J. Theo. Phys. {\bfseries{54}}, 1567 (2015).

\bibitem{mi4199}  K. N. Mishra, D. K. Chakraborty, Astro. Spac. Sci. {\bfseries{260}}, 441 (1999).
\bibitem{te0903}  E. Teo, Gen. Relat. Grav. {\bfseries{35}}, 1909 (2003);
\bibitem{saqib}S. Hussain, I. Hussain, M. Jamil, Eur. Phys. J. C {\bfseries{74}}, 3210 (2014).
\bibitem{ja2415}M. Jamil, S. Hussain, B. Majeed, Eur. Phys. J. C  {\bfseries{75}}, 24 (2015).
\bibitem{fr2012} V. P. Frolov, Phys. Rev. D {\bfseries{85}}, 024020 (2012).
\bibitem{bo2013}C. V. Borm,  M. Spaans, Astron. Astrophys. {\bfseries{553}}, L9
(2013).
\bibitem{mc2307} J. C. Mckinney,   R. Narayan, Mon. Not. Roy. Astron. Soc. {\bfseries{375}}, 523 (2007).
\bibitem{do2008} P. B. Dobbie, Z. Kuncic, G. V. Bicknell and R. Salmeron, Proceedings of IAU Symposium 259
Cosmic Magnetic Field: From Planets, To Stars and Galaxies(Tenerife,
2008).
\bibitem{frolov1998}V. P. Frolov, I. D. Novikov, Black Hole Physics,
Basic Concepts and New Developments, (Springer 1998).
\bibitem{gulmina}G. Z. Babar, M. Jamil,  Y-K. Lim, Int. J. Mod. Phys. D {\bfseries{25}}, 1650024 (2016).
\bibitem{ka1107}  R. Kaya, Gen. Relativ. Grav. {\bfseries{39}}, 211 (2007).
\bibitem{saqib-mog}Hussain, S., Jamil, M.:
Phys. Rev. D {\bfseries{92}}, 043008 (2015).




\bibitem{BSW} M. Banados, J. Silk, S. M West, Phys. Rev. Lett. \textbf{103}, 111102 (2009).
\bibitem{za7110} O. Zaslavskii, JETP Lett. {\bfseries{92}}, 571 (2010).
\bibitem{pu5211} D. Pugliese, H. Quevedo and R. Ruffini, Phys. Rev. D {\bfseries{83}}, 104052
(2011).
 \bibitem{ibrar} I. Hussain, M. Jamil, B. Majeed , Int. J. Theo. Phys. \textbf{54},  1567  (2015).
\bibitem{za4313} A. M. A. Zahrani,  V. P. Frolov, A. A. Shoom, Phys. Rev. D {\bfseries{87}}, 084043 (2013).
    \bibitem{saqib-kerr}S. Hussain, I. Hussain, M. Jamil Eur. Phys. J. C \textbf{74}, 3210 (2014).
        \bibitem{majeedRN} B. Majeed, M. Jamil, S. Hussain,    Adv. High Energy Phys. {\bf 2015},  671259 (2015).

\bibitem{jamilschwarz} M. Jamil, S. Hussain, B. Majeed, Eur. Phys. J. C \textbf{75}, 24 (2015).
\bibitem{f17}R. Saffari and S. Rahvar, Phys. Rev. D {\bf {77}}, 104028 (2008) [arXiv:0708.1482 [astro-ph]].
    \bibitem{azad}A. Qadir, and A. A. Siddiqui, Int. J. Mod. Phys. D {\bfseries{16}}, 25 (2007).
\bibitem{bushra}B. Majeed and Azad A. Siddiqui, Proc. Int. Conf. Relativ. Astrophys., \textbf{129}
(2015), Department of Press and Publications, University of the Punjab,
Lahore.
%\bibitem{hu0815}S. Hussain and M. Jamil, Phys. Rev. D {\bfseries {92}}, 043008 (2015).
%\bibitem{p1} Gal'tsov, D. V. and Petukhov, V. I.: Phys. J. Exp. Theor. Phys. {\bfseries{47}}, 419 (1978).
\bibitem{bsw}O. Zaslavskii, JETP Lett. {\bfseries{92}}, 571 (2010).
%\bibitem{wald8074} Wald, R. M.: Phys. Rev. D {\bfseries{10}}, 1680 (1974).
%\bibitem{al4178}  Aliev, A. N. and Ozdemir, N.: Mon. Not. Roy. Astron. Soc. {\bfseries{336}}, 241 (1978).
\bibitem{frolove3410} V. P. Frolov and A. A. Shoom, Phys. Rev. D {\bfseries{82}}, 084034 (2010).

\end{thebibliography}
  \end{document}